\documentclass[11pts]{article}
\usepackage[dvips]{graphicx,color}
\usepackage{amssymb,amsmath,latexsym,multicol}
\newcommand{\be}{\begin{equation}}
\newcommand{\ee}{\end{equation}}
\newcommand{\ben}{\begin{eqnarray}}
\newcommand{\een}{\end{eqnarray}}
\newcommand{\bc}{\begin{center}}
\newcommand{\ec}{\end{center}}

\begin{document}

{\Large \textbf{A Method to Separate Stochastic and Deterministic Information from Electrocardiograms}}\\

R.M. Guti\'errez\footnote{e-mail: rgutier@uan.edu.co} and L. A. Sandoval\\

{\small
Centro de Investigaciones, Universidad Antonio Nari\~{n}o, Calle 58A No 37-94, Bogot\'a, Colombia.
}\\


\begin{multicols}{2}
\noindent {\large \textbf{Abstract}}\\

{\small \noindent  In this work we present a new idea to develop a method to separate stochastic and deterministic information contained in an electrocardiogram, ECG, which may provide new sources of information with diagnostic purposes. We assume that the ECG has information corresponding to many different processes related with the cardiac activity as well as contamination from different sources related with the measurement procedure and the nature of the observed system itself. The method starts with the application of an improuved archetypal analysis to separate the mentioned stochastic and deterministic information. From the stochastic point of view we analyze Renyi entropies, and with respect to the deterministic perspective we calculate the autocorrelation function and the corresponding correlation time. We show that healthy and pathologic information may be stochastic and/or deterministic, can be identified by different measures and located in different parts of the ECG.} \\

\noindent  {\large \textbf{1. Introduction}}\\

An electrocardiogram, ECG, is a time series of  measurements of  one observable of a complex system: surface electric potentials measured between two poles around the heart. The location of the poles depend on the \emph{derivation} in use [1]. We may consider that the complex system under study is the cardiac activity constituted by several processes, and also related to many other processes including neural, mechanical, hormonal, etc. Therefore, the ECG contains mixed information of different sources and time scales. The nature of the useful information, deterministic or stochastic, that can be extracted from an ECG, depends on the characteristics of the corresponding underlying process, on the process of measurement, and on the capabilities of detecting and differentiating deterministic from stochastic information. The separation of these two kinds of information is difficult, in particular for ECGs and other time series of physiological origin. In general applications, the \textit{a priori} knowledge that we have of the information  and/or the contamination, facilitates this separation, which are not well defined in the case of physiological signals [2,3]. Traditionally, the information contained in an ECG has clinical relevance when it is visually identified by an expert cardiologist [1]. The stochastic and deterministic information extracted from the ECG may provide new sources of information that cannot be identified visually and, therefore, they give complementary information to measure the quality of the cardiac activity using the ECG as the unique source of information not accesible by other means.\\
In this work we present a method to detect and characterize information not visually detectable in the ECG, it can be deterministic or stochastic. We consider that the ECG contains deterministic information, stochastic information and contamination. The first is called deterministic because we assume that in principle there is a deterministic model that may represent such information. The stochastic part can also be described with models of the evolution of the probability distribution of the possible states of the system. The contamination is a consequence of the real limitations of the ECG as a time series: finite resolution, finite number of data points, measurement sources of noise and nonstationarity.\\
The method is based on an improved version of an archetypal analysis by constructing a special base of archetypes to capture relevant characteristics of the ECG. In general, the reconstruction of a signal with archetypes is selfconsistent because the base of archetypes is constructed from the signal itself. What is specific of a particular signal is how the archetypal base is constructed. Recognizing the strong but not perfect periodicity of the ECG, we have prepared a particular archetypal base which permits us to overcome the dominant periodicity of the ECG and thus identify and measure small variations as deterministic and stochastic information. These two informations cannot be detected by visual analysis of the ECG as is traditionally done by a cardiologist. In this work we perform a numerical experiment controlling two known ECGs from a healthy and a pathological case, instead of a large statistical study with many different ECGs, to detect and measure stochastic and deterministic information. It is important to highlight that the construction of the special base of archetypes does not reduce the detected stochastic and deterministic informations to the variability of the R-R intervals. The R-R series is used as an internal system of reference for the archetypes as selfconsistency of the method. One purpose of the numerically controlled experiment is preciselly to show how the changes not relevant to the visual analysis of an expert, such as small dampening of the T wave, may be clearly measured by this method. This provides a new source of information useful for diagnostic and possibly with certain predictive power identifying tendencies of cardiac activity before they become clear pathologies.\\ 
In section 2 we present the procedure to filter out the two sources of  contamination mentioned above corresponding to nonstationarity and noise, and the method to perform the separation of the deterministic and stochastic information. This includes the construction of the archetype-base as will be presented. In section 3 we present the application of the method to the mentioned ECGs in order to characterize the information. Section 4 presents the results and section 5 corresponds to the discussion and conclusions of this work.\\

\noindent {\large \textbf{2. Preprocessing and Archetypal Analysis}}\\

Before we study the ECGs we apply a preprocessing of the signal in order to get rid of some contamination but taking care of the rest of the information contained in the ECG. Given the finite resolution and the finite time of a typical ECG, we consider two sources of contamination that have to be filtered out before any further analysis is performed. We study four ECGs: the first corresponds to a healthy patient 25 years old male, H, 
the second is the same healthy ECG but with its T wave smoothed with a normal local average, SP1; this is
an apparently healthy ECG but we call it Simulated pathology 1 because of its smoothed T wave. The third 
ECG also corresponds to a simulated pathology obtained from the healthy ECG by suppressing the T-wave, SP2; this pathology corresponds to a myocardial damage [1]. The last ECG corresponds to a real pathologic case, MITP, it is the file 100 of the MIT-BIH Arrhythmia Database [4], corresponding to a supraventricular ectopy [1,4].\\
In the extreme of low frequencies, or large time scales, we observe trends or modulations. These trends are filtered out by a time space filtering replacing each data point by the average of itself and its 70 neighbors on each side using a Gaussian distribution. Since the measuring frequency is 300Hz, this average covers 0.47 seconds which corresponds approximately to one half of the time between successive heart beats. However, the Gaussian distribution makes the significative number of neighbors used for smoothing to be 10 on each side of each data point, approximately a time span of 0.07 seconds which does not compromise any relevant structure of the ECG. This procedure also filters out high frequency noise observed as fast fluctuations. In figure 1 we present a few seconds of the four mentioned ECGs that will be studied in this work. We do not make any further treatment to filter out any other contamination considering that the information, deterministic and stochastic, may be distributed in most of the frequency range of the whole power spectrum of the ECG.\\
We first find the R peaks of the ECG. Then we normalize the R-to-R pseudo-periods with a length equal to the greater R-R distance. The traditional archetypal analysis constructs the set of archetypes from the whole set of R-to-R pseudo-periods. We group the R-to-R pseudo-periods in five groups ordering their size from the smallest to the largest. Then we find the set of archetypes for each group. The middle group, the one around the mean value, has most of the RR intervals. We take the first three archetypes obtained from this group. Proportionally, we take two from the two groups at each side of the central group, and one from each of the two external groups. See figure 2. These nine archetypes constitute the archetype-base to reconstruct the corresponding ECG. See figure 3. The archetypes of each group are estimated as follows [5-7]:
Consider a set of multivariate data $\{\textbf{x}_i, i=1,...,n\}$ where each $\textbf{x}_i$ is an $m$-dimensional vector. By means of Archetypal Analysis we search a set of $m$-dimensional vectors $\textbf{z}_j$ that characterize the archetypal patterns in the data. The patterns $\textbf{z}_1,...,\textbf{z}
_p$ are mixtures of the data values $\{\textbf{x}_i\}$. Specifically, let $\textbf{z}_k = \sum_i \beta_{ki} \textbf{x}_i$ be an archetypal element. Here $\beta_{ki} \geq 0$ and $\sum_i \beta_{ki} = 1$. The $\{a_{ik}\}$, $k=1,..,p$, are defined as the minimizer of  $||\textbf{x}_i - \sum_k a_{ik} \textbf{z}_k ||$. Here $a_{ik} \geq 0$ and $\sum_k a_{ik} = 1$. Finally, we define the archetypal patterns as the mixtures $\textbf{z}_1,...,\textbf{z}_p$  that minimize 
$\sum_i ||\textbf{x}_i -\sum_k a_{ik} \textbf{z}_k ||^2$.
\begin{center}
\includegraphics[height=5cm,width=8.2cm]{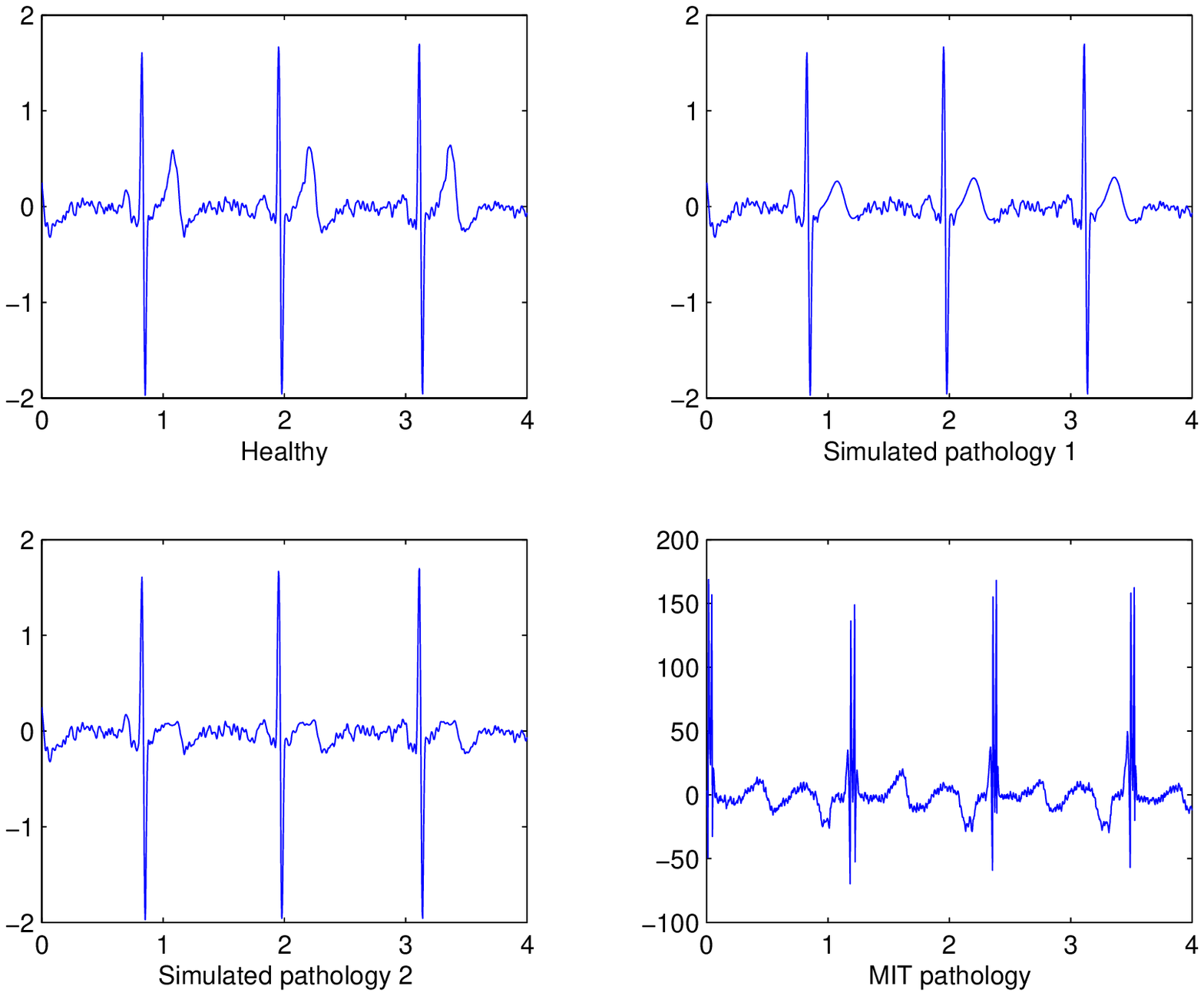}
\end{center}
{\footnotesize Fig. 1 A few seconds of the ECGs H, SP1, SP2  and MITP.}\\

\noindent {\large \textbf{3. Archetypal Coefficients Analysis}}\\

The reconstruction of each ECG from the archetype-base, generates a time series of values for each archetype and each ECG. These values measure the contribution of the corresponding archetypes as they are compared with the succesive RR intervals. We analyze the time series of each coefficient for each ECG to measure the stochastic and deterministic information potentially contained in each case. 
\begin{center}
\includegraphics[height=5cm,width=8.2cm]{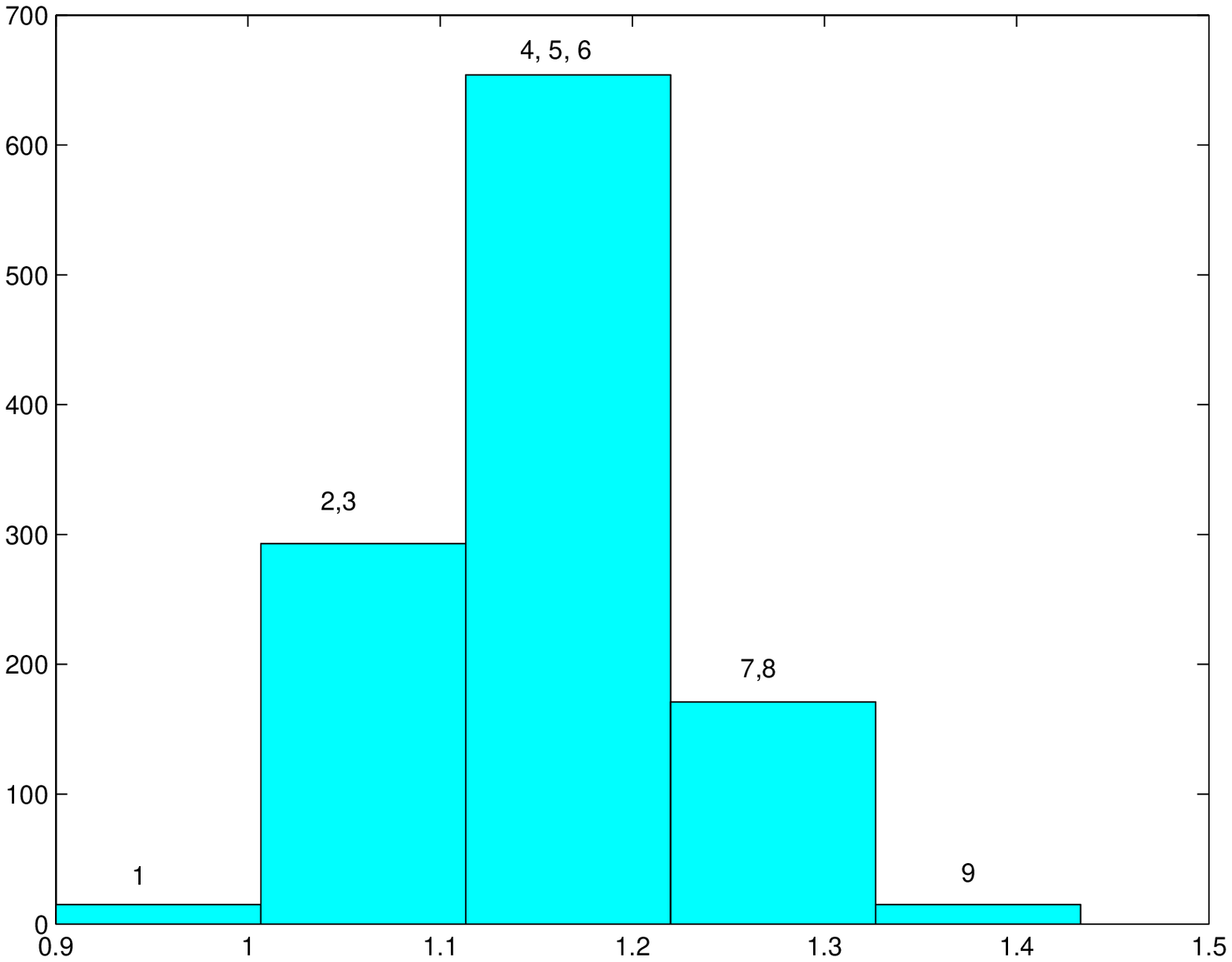}
\end{center}
{\footnotesize Fig. 2 RR interval histogram. The horizontal scale indicates 
the time intervals of the RR series. The numbers on top of the bars indicate the
archetype-base used of 9 archetypes.}\\

We first perform a stochastic analysis of these time series using a symbolic representation. The range 
of values of any coefficient is between 0 and 1. We reduce all the possible values to four symbolic values, 0, 1, 2 and 3, corresponding to the ranges of 0 to 0.25, 0.25 to 0.5, 0.5 to 0.75 and 0.75 to 1 respectively. This approach reduces the details but highlights the most representative qualitative characteristics. With these four symbols we can construct $4^3=64$ words of three symbols, and then we obtain the probability distribution for each time series representing the 9 coefficients obtained from each of the four ECGs analyzed in this work. The Renyi entropies given by 
\be
H^{(q)}_{k} = (1-q)^{-1}log (\sum_{s^{k}EA^{k}} p(s^k)^q)
\ee
give a quantitative measure of the stochastic information contained in these distributions [8]. From the deterministic point of view, we 
calculate the autocorrelation function of these 36 time series of coefficients and estimate the correlation time for each case. The autocorrelation function
\be
C(t) = {1\over N} \sum_{i=1}^{N}{x_i x_{i+t}\over x_{i}^{2}}\,, 
\ee
measures time or causal correlations of the values on a time scale $t$ [9,10]. \\
\begin{center}
\includegraphics[height=7cm,width=8.2cm]{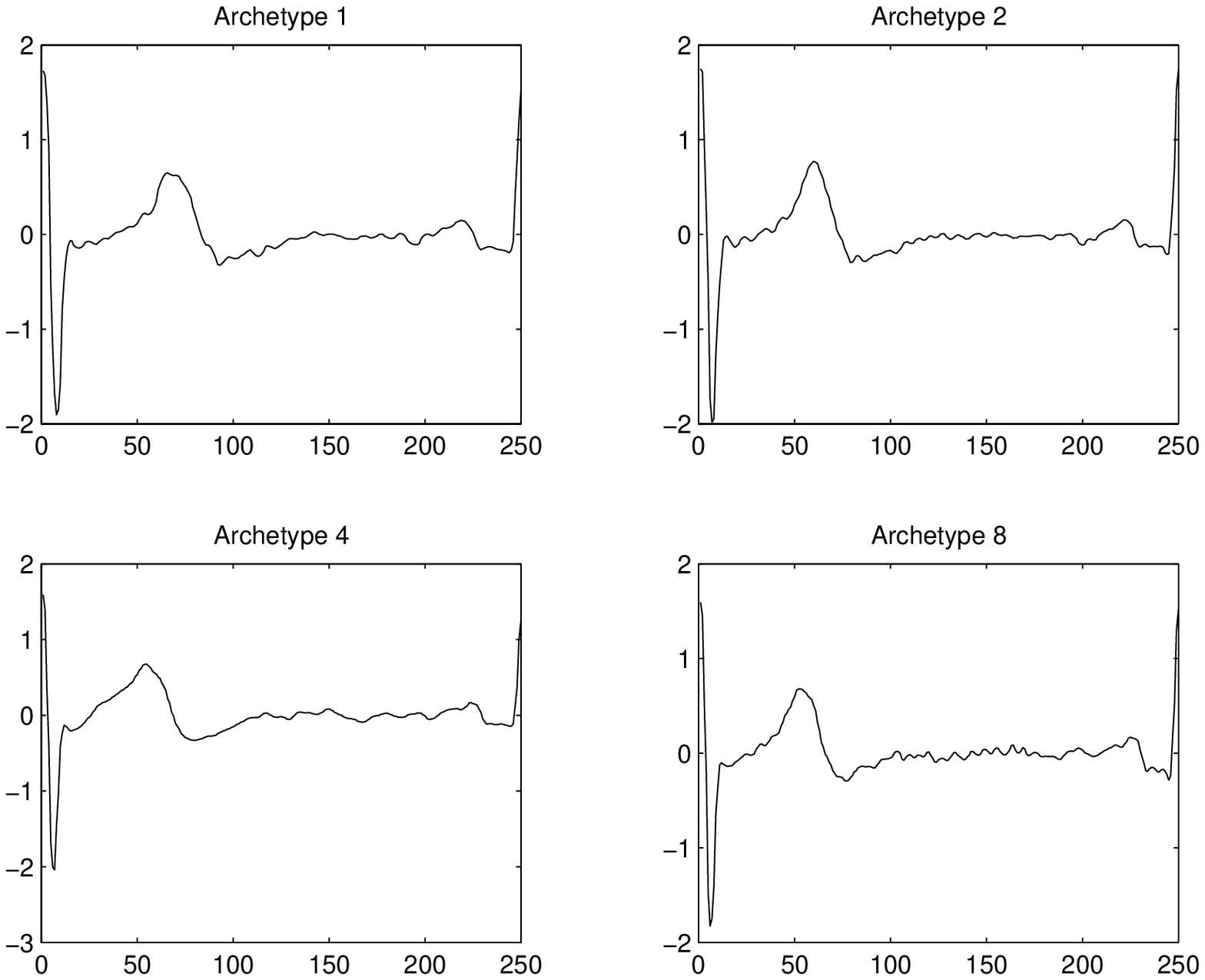}
\end{center}
{\footnotesize Fig.3 Some selected archetypes obtained from the healthy ECG numbered according to figure 2.}\\

\noindent {\large \textbf{4. Results}}\\

In figures 4 to 6 we present the most interesting results of this work. In figure 4 we plot the average
of the archetypal coefficients for the four ECGs as indicated in the figure. These averages are the mean 
values of each coefficient. A coefficient mean value is obtained from the values of the coefficient as 
the corresponding archetype is compared with each RR interval. 
\begin{center}
\includegraphics[height=7cm,width=8.2cm]{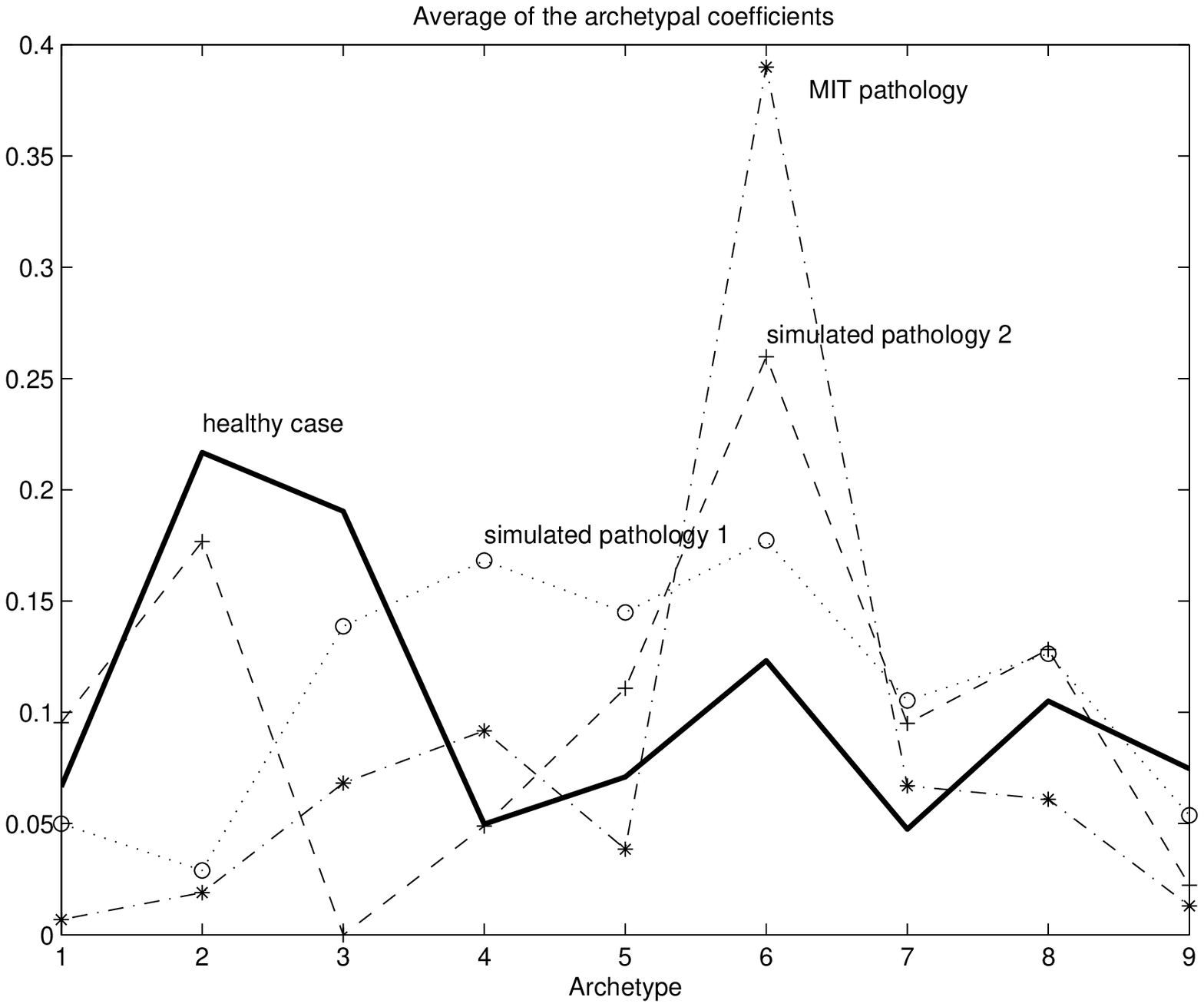}
\end{center}
{\footnotesize Fig. 4 Averages of the archetypal coefficients for the four ECGs.}\\

This procedure measures the contribution 
of the corresponding archetype to the morphology of the RR interval. Therefore, the mean value of the coefficient of
each archetype measures the importance of this archetype for the morphology of the whole ECG.\\
\begin{center}
\includegraphics[height=6cm,width=8.2cm]{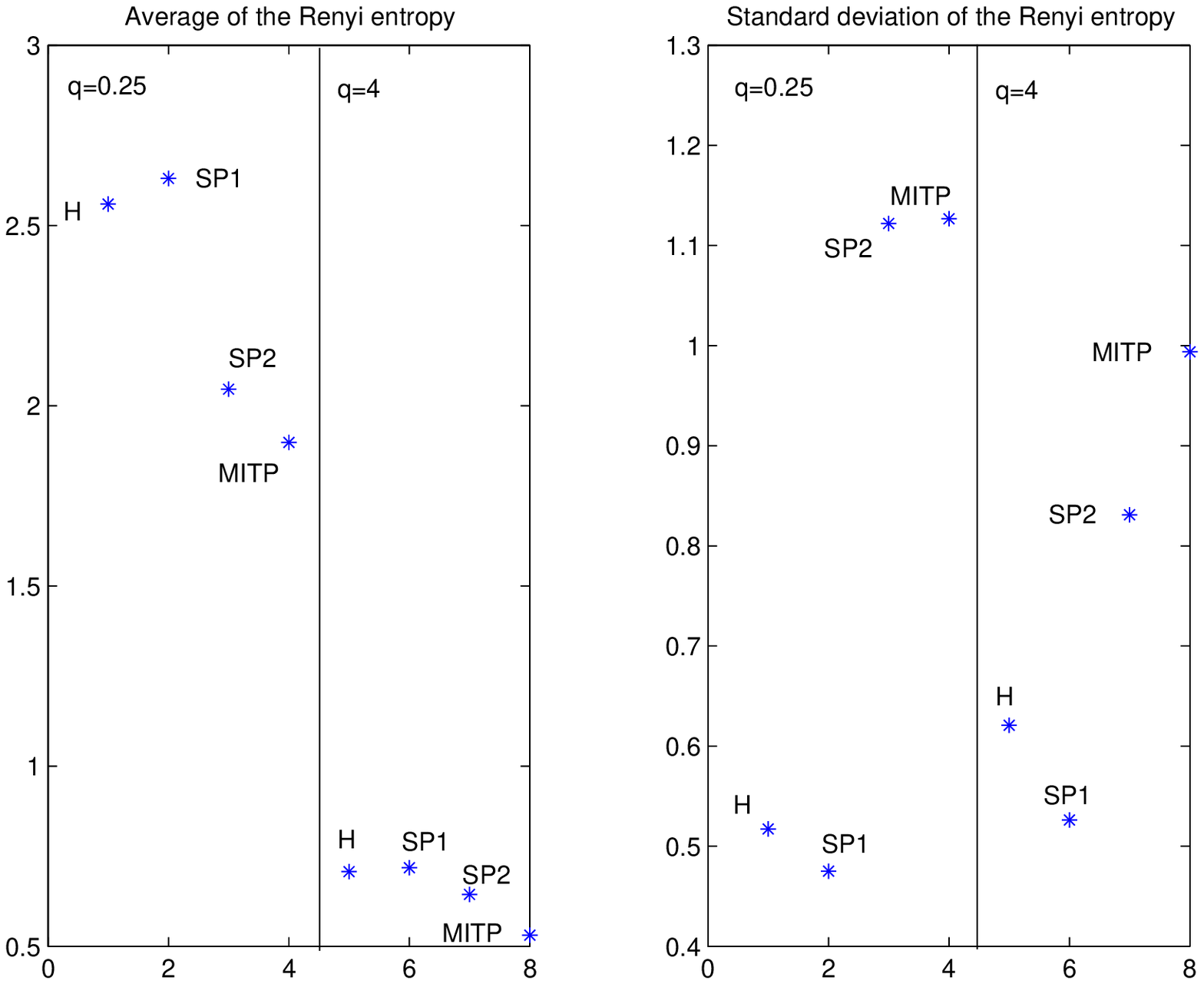}
\end{center}
{\footnotesize Fig. 5 Average Renyi entropies and their standard deviations for $q=4$ and $q=1/4$ for the 4 ECGs as indicated.}\\

Figure 5 presents the results of two different Renyi entropies, for $q=1/4$ and $q=4$. These entropies are 
measures of disorder where the small and large probabilities dominate respectively. The healthy and T-wave 
smoothed ECG, H and SP1, present larger average entropies and smaller standard deviations for $q=1/4$ than the two 
artificial and real pathologic ECGs, SP2 and MITP. These results indicate that the stochastic information is 
homogeneously distributed in healthy ECGs and that this distrubution of stochastic information is lost
in the pathologic cases.\\
\begin{center}
\includegraphics[height=7cm,width=8.2cm]{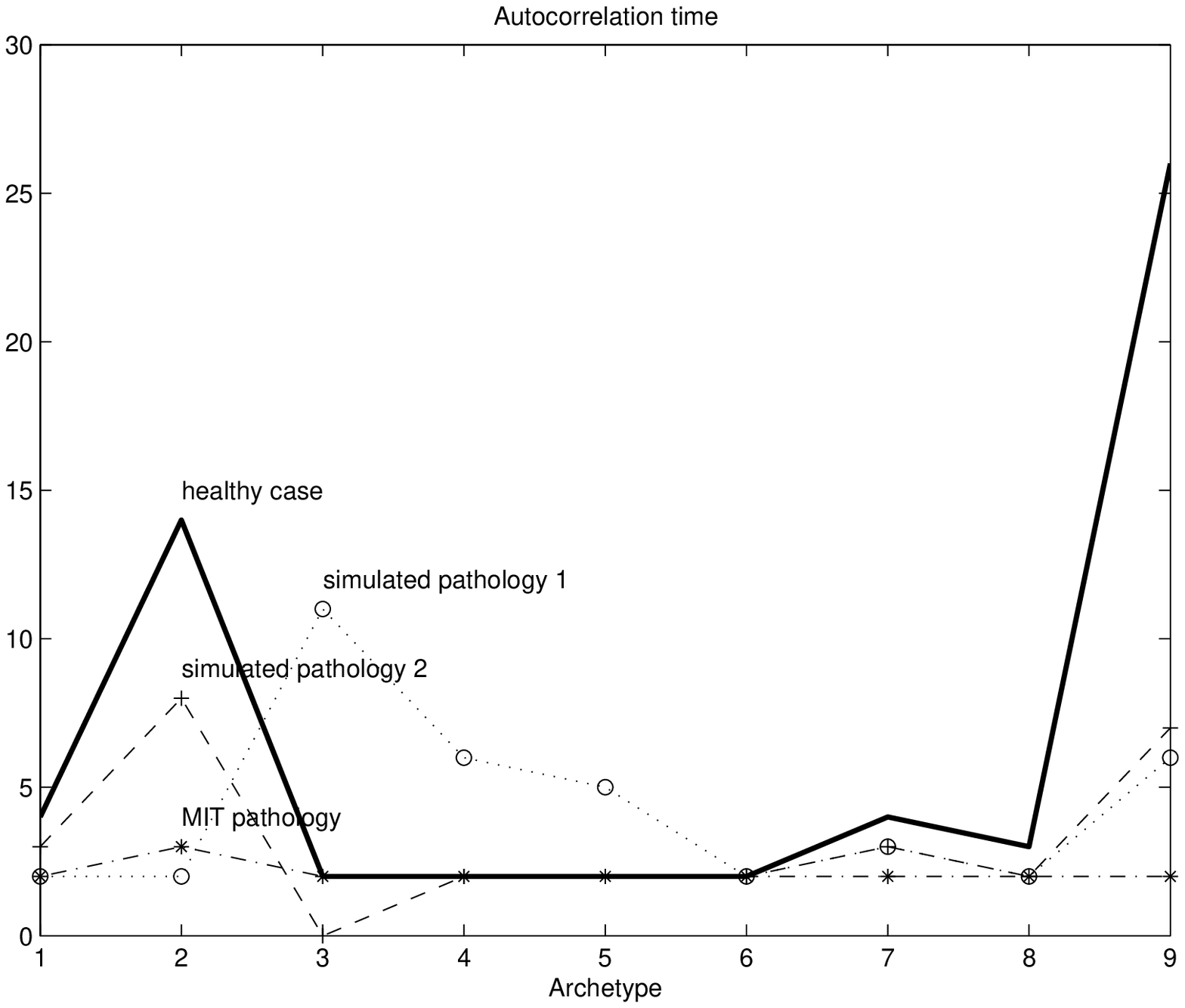}
\end{center}
{\footnotesize Fig. 6 Autocorrelation times for the 9 coefficients of each archetype and the four ECGs as indicated.}\\

Figure 6 shows the correlation time $t$, for the nine archetype coefficients of the four ECGs, estimated from equation (4) when $C(t)=1/3C(0)$. We observe that correlation times are larger for healthier ECGs, H and SP1, and smaller for pathologic ECGs, SP2 and MITP. We can also observe appreciable differences between H and SP1, and SP2 and MITP.\\

\noindent {\large \textbf{5. Discussion and Conclusions}}\\

With this work we do not pretend to find new physics nor develop all the technical details and standarization of a new tool for diagnostic using the ECG as the unique source of information. As we understand the scope of applied physics, we present a method of applying new concepts with improved techniques and show its potential to help in the solution of actual problems in cardiac diagnostics. For the discussion of results we have to keep in mind that SP1 actually resembles a healthy case, whereas SP2 does not.\\
If a given coefficient does not change apreciably along the analyzed ECG, it means that the morphology represented by this particular archetype is invariant throughout the ECG. On the other extreme, if the values change a lot and randomly, the corresponding morphology changes randomly. An intermediate behavior may indicate some order that can be characterized by a stochastic distribution and/or deterministic correlations. \\
The mean values of $H^{1/4}$ for H are large and a little larger for SP1. The corresponding values of $H^{1/4}$ for the SP2 are small and still a little smaller for MITP. This means that the morphologies represented by the archetypes are, in average, more disordered for the healthy ECG than the corresponding to the two, artificial and real, pathologic cases, SP2 and MITP. In addition the standard deviations of $H^{1/4}$ are smaller for H and SP1, and larger for SP2 and MITP. This indicates that the morphologies represented by the archetype are homogeneously distributed over the nine archetypes of the archetype-base; for the pathologic cases the disorder is more localized in some archetypes than in others. Therefore, healthier ECGs present more disorder, higher entropies, than the pathologic ECGs. \\ 
The correlation time, obtained from the autocorrelation function of the coefficients, shows that the local morphologies of the healthier ECGs are more correlated than the pathologic ECGs. This temporal correlation indicates some deterministic information in the ECG that is more evident in healthier ECGs. As a conclusion, we observe that the method presented in this work detect strong evidences of stochastic and deterministic useful information in ECGs.  These two kinds of information can diferentiate pathologic and healthy ECGs even when the differences cannot be detected by visual analysis.  We have succeeded to some extent, to separate these two informations from the ECG, measure them for some trial controlled and known cases, and show the distinctic characteristics of each one as they are extracted from the different ECGs. These results are consisten with the complex nature of the cardiac dynamics where stochastic and deterministic aspects are both present in a complex mixture. The quality of cardiac dynamics may be characterized from the ECG as the unique source of information.\\

\noindent {\large \textbf{References}}\\
{\small
\begin{enumerate}
\item H. V\'elez, W. Rojas, J. Borrero and J. Restrepo, ``Fundamentos de Medicina, Cardiolog\'ia. Corporaci\'on para la Investigaciones Biol\'ogicas. Colombia, (1997); A. C. Gyton and J. E. Hall, ``Textbook of Medical Physiology", 9th edition, W. B. Saunders Co., Phyladelphia, (1997).

\item A. Eke, P. Herman, L. Kocsis and L. R. Kozak. ``Fractal characterization of complexity in temporal physiological signals". Physiol. Meas. 23 R1-R38, (2003).

\item R. M. Guti\'{e}rrez and L. Sandoval. ``Detecting the Stochastic and Deterministic Information of ECGs". Proceeding of The 6th World Multiconference on Systemics, Cybernetics and Informatics. Orlando, Florida. Nagib Callaos ed., v.II. p. 290, (2002).

\item http://www.physionet.org/

\item A. Cutler and L. Breiman, ``Archetypal Analysis'', TECHNOMETRICS, V 36, NO. 4, p. 338, (1994). 

\item M. D. Ortigueira et al., ``An archetypal based ECG analysis system''. III Congress of Matlab Users, MatLab'99, Madrid, Spain, 17-19, p. 467, (1999).

\item M. D. Ortigueira, ``Archetypal ECG Analysis'', Proceedings of RECPAD-98, Instituto Superior T\'ecnico, Lisboa, Portugal, p. 373, (1998).

\item J. Kurths et al., ``Measures of complexity in signal analysis''. Proceeding 3th Technical Conference on Nonlinear Dynamics (chaos) and Full Spectrum, New London, July 10-13, (1995).

\item H. Kantz and T. Schreiber, ``Nonlinear Time Series Analysis". Cambridge University Press, Cambridge (1997).

\item J. Kurths et al., `` Quantitative Analysis of Heart Rate Variability". Chaos 5, 88 (1995).

\end{enumerate}
}
\end{multicols}
\end{document}